# Record High-Gradient SRF Beam Acceleration at Fermilab

D.Broemmelsiek, B.Chase, D.Edstrom, E.Harms, J.Leibfritz, S.Nagaitsev, Y.Pischalnikov, A.Romanov, J.Ruan, W.Schappert, V. Shiltsev* , R.Thurman-Keup, A. Valishev

Fermi National Accelerator Laboratory, PO Box 500, Batavia, IL 60510, USA

*e-mail: shiltsev@fnal.gov

**Abstract**

Many modern and future particle accelerators employ high gradient superconducting RF (SRF) to generate beams of high energy, high intensity and high brightness for research in high energy and nuclear physics, basic energy sciences, etc. In this paper we report the record performance large-scale SRF system with average beam accelerating gradient matching the ILC specification of 31.5MV/m. Design of the eight cavity 1.3 GHz SRF cryomodule, its performance without the beam and results of the system commissioning with high intensity electron beam at FAST (Fermilab Accelerator Science and Technology) facility are presented. We also briefly discuss opportunities for further beam studies and tests at FAST including those on even higher gradient and more efficient SRF acceleration

## 1. Introduction

Over the past three decades, since pioneering realizations at CEBAF at JLab, TRISTAN at KEK and LEP-II at CERN in 1980s, the science and technology of superconducting radio-frequency (SRF) beam acceleration have matured and tremendously advanced. Widespread employment of SRF to improve the performance of existing accelerators and construction of new facilities is mostly due to such advantages of the technology as extremely low RF losses in the cavity walls at cryogenic temperatures, high wall plug to beam power conversion efficiency, possibility of long beam pulse operating modes, reduced interaction of the beam with larger aperture cavities leading to low impedance for high current beams, and high accelerating gradient, $E_{acc}$ [1-3]. Table 1 presents key parameters for major linear accelerators based on the pulsed SRF, such as the Spallation Neutron Source at ORNL [4], European Spallation Source in Sweden [5], the "Proton Improvement Plan – II" linac at Fermilab [6], European X-FEL at DESY [7], MARIE FEL at LANL [8] and the International Linear Collider in Japan [9]. The SRF system of a particle accelerator consists of many subsystems and components, such as RF cavities, fundamental RF input power couplers, high-order modes (HOMs) dampers, frequency tuners, cryostats, high- and low-level RF systems, instrumentation, vacuum system, and cryogenics, and all that should be coupled to the rest of the accelerator complex (beam sources and absorbers, beam diagnostics and control, vacuum, collimation and radiation protection, safety, tunnel environment control and interlocks, etc). Because of the additional complexity, beam operations or realistic tests with beams are considered ultimate indicators of achievement. As an example, the SNS linac beam operations has shown some degradation of the SRF cavity gradients which are routinely at the level about 13MV/m on average, that is below the design specification (though some cavities indeed are at or even above the 15.8 MV/m level) [4].

The ILC is one of the most complex accelerator projects ever considered and designed [9]. Each of the two ILC linacs will accelerate electrons or positrons from 15 GeV to 250 GeV through 7332 9-cell 1.3 GHz SRF cavities, that is about 32 MeV per cavity. The cavities operate at 2 °K and are grouped into approximately 12.6 m long cryomodules of two types with either 8 or 9 cavities in them, and therefore, the beam energy gain in the 8 cavity cryomodule needs to be 32·8=256 MeV. To meet such specification with 1.5% energy overhead, each 1.038m long cavity should on average have 31.5 MV/m accelerating gradient with operationally acceptable spread of ±20%. It is of utmost importance to provide a highly reproducible and stable beam and a powerful and flexible control system is needed for precision regulation of the RF fields inside the accelerating cavities.

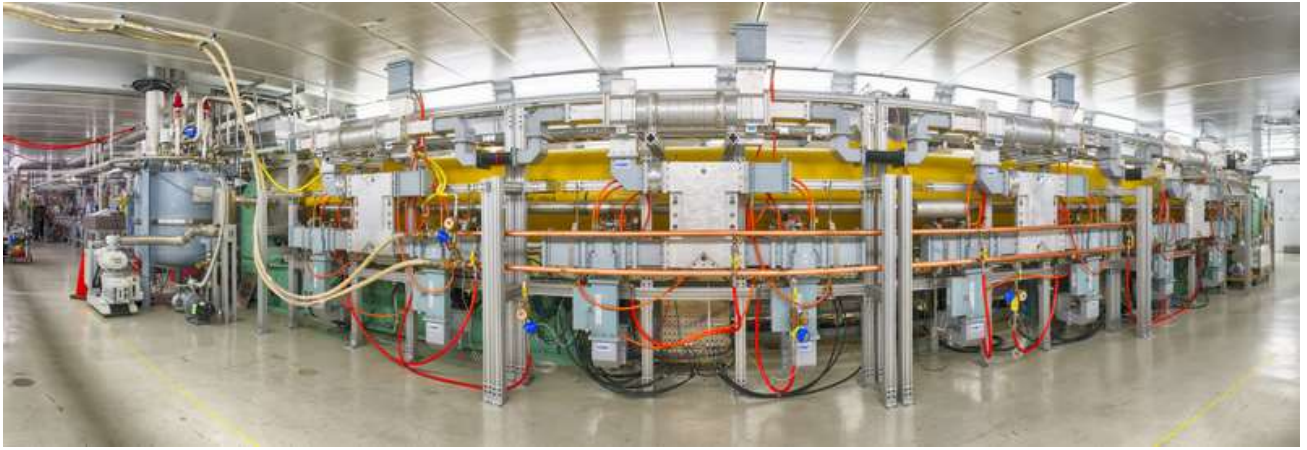

Figure 1: The 1.3 GHz SRF cryomodule (CM2) installed in the FAST/IOTA facility (courtesy: Fermilab).

The so-called "S1-Global collaborative project" [10] was set up by the ILC Global Design Group to demonstrate achievement of a 31.5 MV/m gradient in a cavity string in one 1.3 GHz SRF cryomodule and perform a realistic system test with high current beam acceleration. Even getting the required average gradient without a beam, with RF only in a single cryomodule test stand was found to be formidable task [11], which was finally achieved at Fermilab [12]. Beam acceleration by the ILC/TESLA-type 1.3GHz SRF cryomodules was first demonstrated at the FLASH facility at DESY with average gain per eight cell CM of 150-180 MeV initially and 200-240 MeV with an X-FEL type CM [13], still below the ILC specification. Recent beam commissioning of the European X-FEL has shown that a few of about 100 cryomodules can achieve 27.2 MV/m gradients [14], well above the X-FEL specification but short of the ILC goal of 256 MeV per SRF cryomodule. In this paper, we report for the first time the record high gradient SRF beam acceleration which meets the ILC requirement.

Table 1: Design parameters of large pulsed SRF accelerator facilities.

| Facility | Accelerating gradient, MV/m | RF frequency, MHz | Repetition rate, Hz | Beam pulse length, ms |
|---|---|---|---|---|
| SNS | 15.8 | 805 ($\beta$=0.81) | 60 | 0.7 |
| ESS | 19.9 | 704 ($\beta$=0.86) | 14 | 2.86 |
| PIP-II | 18.8 | 650 ($\beta$=0.92) | 20 | 0.54 |
| X-FEL | 23.6 | 1300 | 10 | 0.65 |
| MARIE | 31.5 | 1300 | 10 | 0.1 |
| ILC | 31.5 | 1300 | 5 | 0.73 |

Fermilab has built an ILC-type cryomodule RFCA002 (which will be marked as "CM2" below) which is comprised of eight 1.3GHz cavities. The cavities were tested in vertical and horizontal test stands at Jefferson Lab and Fermilab and all reached gradients in the range 33-41 MV/m before quenching [12]. The assembled CM2 was installed at Fermilab Accelerator Science and Technology (FAST) facility [16] which houses the necessary cryogenic and RF infrastructure for high power and beam testing of the CM2 at 2 °K– see Fig.1.

FAST consists of the Integrable Optics Test Accelerator (IOTA) ring, capable of operating with 150-300 MeV/c electrons and 70 MeV/c protons and its two injectors which includes the 1.3 GHz SRF electron linac – see Fig.2. The facility is being built to host the US-leading accelerator R&D program towards multi-MW beams and experimental tests of novel techniques for high-current beam accelerators. Beam commissioning of the CM2 cryomodule and attainment of the ILC beam acceleration gradients took place in October-November 2017.

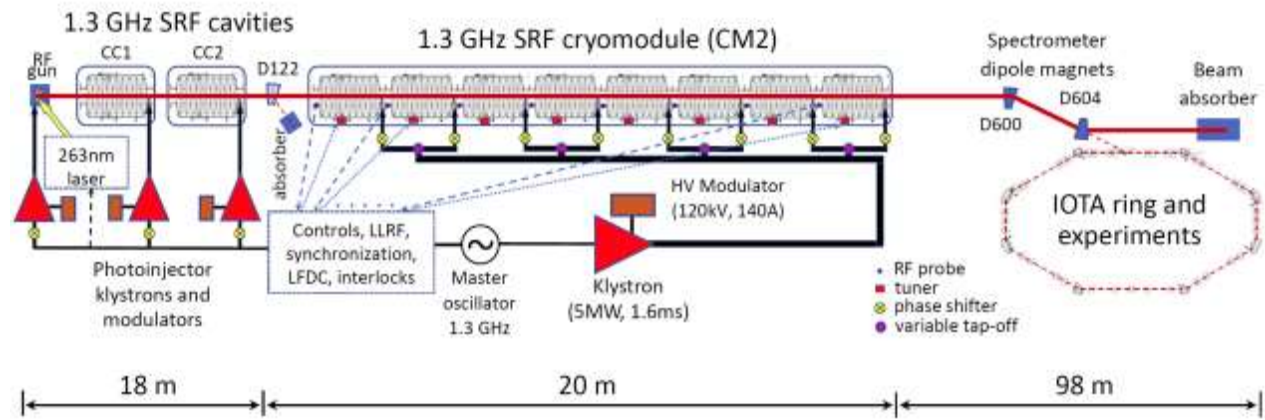

Figure 2: Schematic layout of the FAST/IOTA facility (not in scale).

In Section 2 of this article we briefly present the design of the eight cavity 1.3 GHz SRF cryomodule, its performance without the beam and novel technologies developed to stabilize the frequency, amplitude and phase of each cavity within tight specifications needed for the ILC-type beam acceleration. Results of the system commissioning with high intensity electron beam at FAST are presented in Section 3. Finally, we discuss further beam studies of efficient beam SRF acceleration and opportunities to achieve even higher beam accelerating gradients.

## 2. SRF Cryomodule: Design and Performance

*2.1 SRF Cavities*

The 1.3-GHz SRF accelerating cavities were originally developed in the context of the TESLA linear-collider project [17] and were included in the baseline design of the international linear collider (ILC) [18]. Such a cavity - see Fig.3 - is a 9-cell standing-wave accelerating structure operating in the $TM_{010;\pi}$ mode. Total cavity flange-to-flange length is 1.247m, while active length (weighted with electric field) is 1.038m. The cavity operates at 2°K with design average acceleration gradient $E_{acc}$=31.5MV/m and quality factor $Q_0 \geq 0.8\cdot 10^{10}$. Each cavity has two end group sections - one with a port for coupling RF power from the power source into the structure, and the other with a port for a field sampling probe used to determine and control the accelerating gradient. Each of these ports accepts an electric field antenna required for qualification and operation. In the process of building a cryomodule, these cavity structures are cleaned/processed, tested and "dressed" - welded in a helium jacket for cooling together with additional peripheral components assembled on them, such as coarse and fine tuners for adjusting the frequency of the structure, magnetic shielding material to minimize the cavity losses, a variable coupling high power input antenna for powering the cavity, an electric field sampling antenna and two higher order mode (HOM) electric field antennas. Eight of these dressed cavities are connected into a string and are a subcomponent of a 12.6 m long superconducting cryomodule.

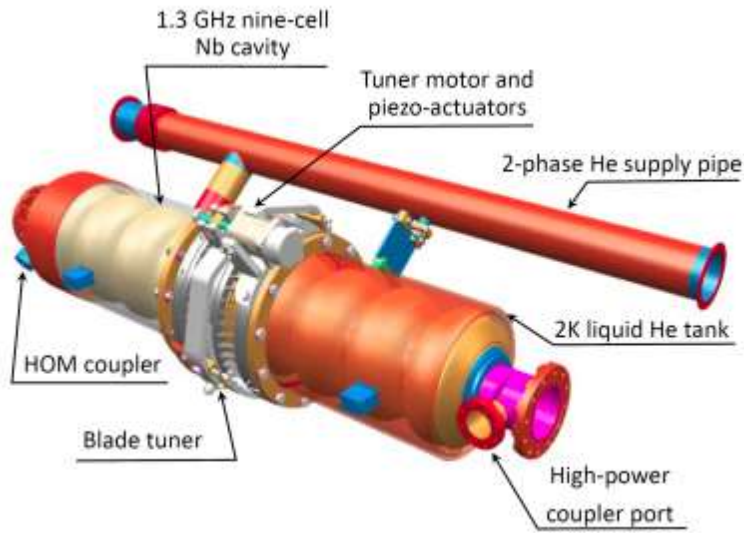

Figure 3: The baseline ILC cavity package ("dressed cavity"), titanium helium tank is shown as transparent for a view of the 9-cell niobium RF cavity inside (adapted from [9]).

Each cavity is equipped with a so called "blade tuner" system that transforms azimuthal rotation into a longitudinal motion and allows changing the length of a cavity coaxial with it. The tuner allows both slow mechanical adjustment of the frequency of the cavity to bring it on resonance (static tuning) and a fast 'pulsed' adjustment using a piezo system to dynamically compensate Lorentz-force detuning during the RF pulse. The tuning range of the cavity frequency is of order 600 Hz.

*2.2 SRF Cryomodule*

The SRF cryomodule CM2 installed in the Fermilab Accelerator Science and Technology (FAST) facility is the second of two ILC style 8-cavity cryomodules built at Fermilab. The cryogenic infrastructure of the facility allows 2°K superfluid Helium cooling of the cryomodule [18]. All cavities were fabricated in industry and extensively tested cold prior to assembly into the cryomodule. While the first cryomodule was assembled by Fermilab staff with oversight by collaborators, parts were provided as a "kit" by DESY, CM2 was largely assembled by Fermilab with parts also obtained by Fermilab. A single 5 MW peak 1.3 GHz klystron provides the necessary power to drive the cryomodule. A waveguide distribution system provided by SLAC National Accelerator Laboratory [19] is attached to the cryomodule fundamental power couplers and splits the incoming power amongst the eight cavities. The low-level RF (LLRF) drive system was developed based on previous systems for SRF test beds and includes provision for both amplitude and phase feedback control within the ILC specifications. Figure 4 compares the gradient performance of all cavities in their various test schemes. Vertical and horizontal tests are conducted on bare single cavities first and then after dressing respectively. At FAST singe cavities tests were first carried out on each cavity alone being driven by the RF source and finally as a unit. Single cavity testing allowed the entire available RF power to be delivered to a single cavity while 'unit' tests necessarily limit the peak RF power available to a single cavity to a maximum of 1/8 of the total output of the klystron.

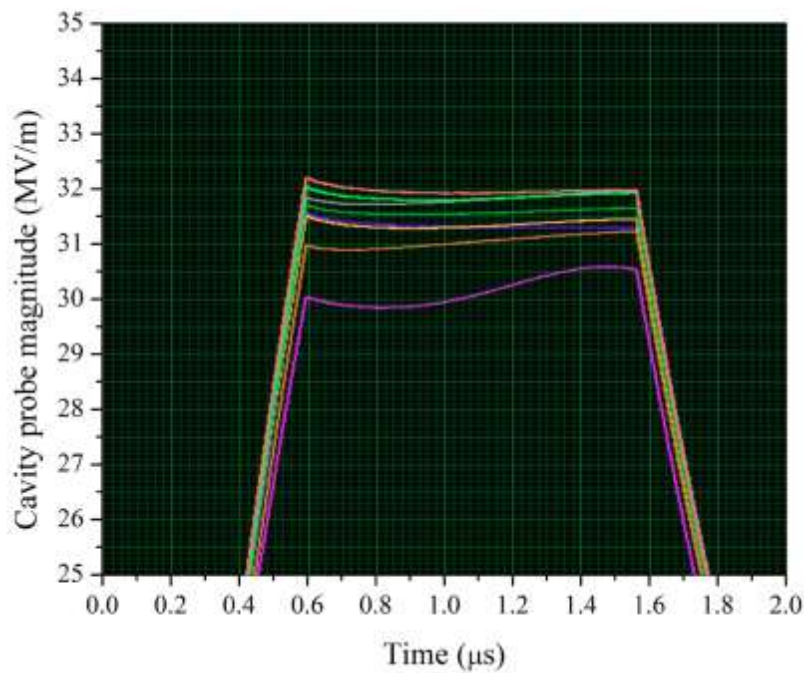

Figure 4: Measured maximum accelerating gradients in all eight SRF cavities of CM-2. The RF pulse has 0.6 ms filling time and 1.0 ms flat-top.

CM2 has now been operated for two extended operations periods – initial commissioning (without beam) from June 2013 through August 2015 and most recently re-commissioning followed by beam operation beginning in July 2017 through November 2017. The main activities and accomplishments of the first run are already described in Refs. [12, 20].

Table 2 summarizes the peak gradients achieved and measured cavities' quality factors $Q_0$ at the design gradient, 5 Hz repetition rate, temperature of 2°K and 1.57 ms long RF pulse length with

0.97 ms flat-top. Most of the resonators are demonstrated to operate above 32 MV/m and are limited at that point either by quench or available RF power. Cavity number 6 is the limiter and can operate only below 30 MV/m before quenching. This limit was demonstrated during both the first commissioning run and verified during the most recent beam run. Since all eight cavities are now driven by a single RF source, single performance is limited by the available power and the amount going to the 'weakest' cavity. As conditions permit re-orienting the RF distribution system would conceivably allow for higher performance of cavities based on their most recent demonstrated performance. At FAST the peak gradient was administratively limited to 31.5 MV/m during initial single cavity assessment in order to avoid any possible damage prior to all cavities reaching this level. This limit has since been raised to 35 MV/m and the gradients in all the cavities but number 6 were limited by the available RF power. Though it was not critical for the CM2 operation at FAST, it is noteworthy that majority of the cavities had measured quality factors $Q_0$ above the ILC TDR specification of $0.8 \cdot 10^{10}$.

Table 2: Peak CM2 cavity gradients and quality factors. Total available accelerating voltage is 267.9 MV, and the average gradient of 32.2 MV/m (2015 measurements). Gradient measurements uncertainty is 0.5 MV/m, accuracy of the quality factor measurements is better than 10%.

| Cavity No. | $E_{acc}$ (MV/m) 2013-15 | $E_{acc}$ (MV/m) 2017 | $Q_0/10^{10}$ at 31.5MV/m |
|---|---|---|---|
| 1 | 32.3 | 31.7 | 0.87 |
| 2 | 31.8 | 31.7 | 1.41 |
| 3 | $\geq 33$ | 31.6 | 1.25 |
| 4 | $\geq 33$ | 31 | 0.88 |
| 5 | $\geq 33$ | 31.7 | 0.85 |
| 6 | 29.8 | 30.5 | 0.64 |
| 7 | $\geq 33$ | 32.3 | 0.57 |
| 8 | 32.2 | 31.2 | 0.76 |
| Sum voltage (MV) | 267.9 | 261.2 | |

*2.3 CM2 Performance during beam tests*

During the most recent test run in 2017 the commissioning protocol established previously was used albeit abbreviated; the klystron output remained oriented to drive all cavities simultaneously. This included a period of warm coupler conditioning followed by cooldown, moving the cavities on resonance, on-resonance conditioning, power calibrations, and finally peak gradient determination. The LLRF system was re-commissioned once cool down was completed and employed to move cavities on resonance. Feedback operation was re-established and gains were found to be virtually unchanged from previous operation. Similarly, adaptive feedforward compensation was rapidly re-deployed with minimal calibration time.

Perhaps not surprisingly, CM2 performance was found to be little changed even after a nearly 2-year hiatus at room temperature. Peak gradients, particularly that of cavity 6, were found to be essentially identical within normal error to that of before. The suite of radiation field emission detectors installed for initial commissioning was no longer available, but the remaining subset show indications of the same behavior as before. $Q_0$ measurements have not been repeated.

In order to achieve the desired accelerating voltage despite cavity number 6's limitation, a few techniques were undertaken to maximize the sum voltage. These steps were done to cavity 6 in order to maintain its operating gradient below its quench limit while keeping the total RF voltage constant and included: slightly lowering its loaded quality factor $Q_L$, from $3.5 \cdot 10^6$ to $3.1 \cdot 10^6$ and de-tuning its resonant frequency by about 50 Hz off the drive frequency $f$=1.3GHz. Collectively the duty factor was lowered from 5 Hz to 1 Hz and the flattop length reduced from the nominal 590 μs to 100 μs in order to reduce the average power/heating. Combined these steps allowed sufficient additional voltage to achieve the total CM2 electron energy gain above 250 MeV.

All in all, performance of CM2 during FAST beam operation has been quite reliable. Some activity in the RF power couplers – field emission and vacuum – has been observed, but have not greatly affected beam stability.

*2.4 Cavity control and stabilization: Lorentz force detuning compensation and microphonics feedback*

The walls of the ILC SRF cavities are 2.8 mm thick to allow effective cooling by Helium but the thin walls make the cavities susceptible to micron-scale mechanical deformations induced by force of the accelerating electromagnetic field on the cavity walls as well as by fluctuations in the pressure of the surrounding helium bath and microphonics induced by external noises sources. These mechanical deformations can change the resonant frequency of the cavity by $\Delta f$ and lead to fluctuations of the phase and amplitude of the accelerating field beyond the ILC specifications of 0.35º and 0.07%, correspondingly [9]. For high-gradient, pulsed cavities operating in super-fluid helium, the dominant effect is the Lorentz force detuning (LFD) due to the radiation pressure of the electromagnetic field during the RF pulse which deforms the cavity and pulls it off resonance proportionally to the square of accelerating field $E_{acc}$. For the ILC cavities the static detuning is approximately $\Delta f \sim - E^2_{acc} \times 1.0$ Hz/(MV/m)$^2$ [21] but response to the pulsed force is complicated by excitation of mechanical vibrations, e.g, of the main resonant mode with about 200 Hz frequency. A reliable automated system to compensate for the LFD is needed, otherwise the ILC cavities would dynamically detune by several times the cavity's bandwidth of $f/(2Q_L)$~200Hz and significant excess RF power would be required to maintain the accelerating gradient at the design level of 31.5MV/m.

The use of piezo actuators to compensate for LFD was pioneered at DESY but has since been adopted widely [22]. The DESY team employed the technique of exciting the piezo actuator with a half cycle of sine wave prior to the arrival of the RF pulse. The duration, delay and amplitude of the half sine wave are optimized manually by trial and error. This method became "standard" and used by many groups [23]. While the standard approach can provide acceptable compensation for LFD, the mechanical response of individual cavities to the Lorentz force and to the piezo actuator can differ. Changes in cavity operating conditions, for example the changes in the gradient or bath pressure can require corresponding changes in the compensating waveform. Operating multiple cavities for extended periods will require control systems that can automatically determine the best parameters for each cavity and adapt to changing operating conditions. Because the cavity detuning does not respond linearly to the changes in some parameters of the standard unipolar pulse, the adaptive capability that can be incorporated in LFD systems based on this "standard" approach may be limited.

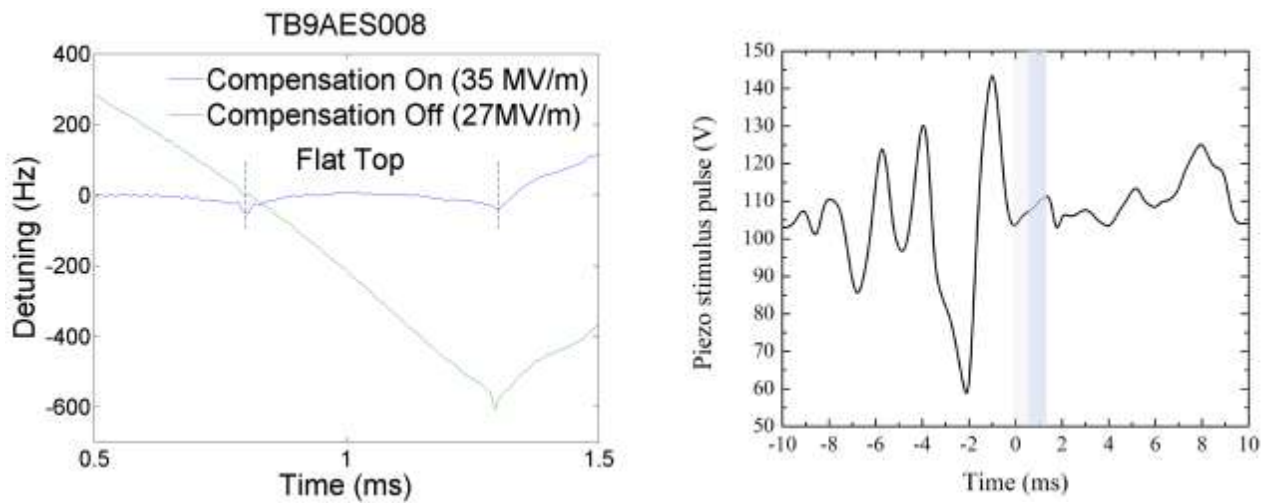

Figure 5: a) left - effective Lorentz force detuning compensation in a nine-cell elliptical cavity number 1 over 0.5 ms flat-top at 35 MV/m; b) right - compensating piezo actuator waveform for the 1.3 GHz SRF cavity operating at $E_{acc}$=35MV/m. Lightly shaded areas indicate the RF filling time period and the pulse flat-top depicted in Fig.5a).

The Fermilab team developed an innovative method to automatically determine an optimal piezo actuator waveform for each individual cavity based on an adaptive feed-forward algorithm [24]. As the first step, the cavity reaction to the piezo impulse is measured at a gradient of several MV/m below the maximum [25]. A series of low-amplitude unipolar 0.5ms long drive pulses are sent to the actuator. The first pulse is timed to arrive some 10 ms in advance of the RF pulse. For each subsequent RF pulse, the delay between the piezo drive and the RF pulse is reduced by 0.5 ms in until the piezo pulse follows the arrival of the RF pulse. The forward, probe and reflected RF waveforms are recorded at each delay and used to determine the detuning. The least squares algorithm is them used to find the combination of the delayed pulses that minimizes the frequency detuning from this data. Due to non-linear response of the system, the response needs to be measured over the range of the bias voltages on the piezo actuator. So, the voltage is stepped from pulse to pulse in small increments over the full range of the actuator and again the RF waveforms are recorded for each pulse. Finally, to measure the detuning due to the Lorentz force alone, the RF waveforms are recorded while no drive signal is sent to the piezo. As a final step the response matrix used to calculate waveform (including bias) that will be applied to piezo for each cavity. As operating conditions vary, the RF waveforms can be used to measure any residual detuning. The same response matrix can then be used to calculate the incremental waveform required to cancel that residual detuning.

Each dressed cavity, equipped with tuner, that was integrated into CM2, has been tested first at Horizontal Test Stand (HTS) at maximum operating gradient in the range of 30 to 35MV/m [26, 27]. A single-cavity LFD compensation system based on the Fermilab adaptive algorithm has been deployed and actively used during the HTS testing [28]. Similar system extended to operate with 8 cavities was successfully deployed to compensate the LFD in the CM2 cryomodule during its commissioning and, later, beam operation [29]. Figure 5 a) shows the detuning with and without compensation of a nine-cell elliptical cavity (TB9AES008) at the HTS during qualification test. Because of limitation in the power of the HTS klystron without LFD compensation cavity cannot be operated above $E_{acc}$=27MV/m. The amplitude of the cavity's LFD during 0.5ms RF pulse "flat top" was almost 600Hz. It is meant that at 35MV/m LFD without compensation could reach ~1000Hz. After turning on the adaptive LFD compensation system it was possible to raise the gradient on the cavity up to 35MV/m. The LFD compensation system suppressed the detuning to less than ±20Hz during whole length of RF pulse. Figure 5b) shows the piezo drive waveform used to compensate the LFD on this cavity.

Cavities can also be detuned by variations in the pressure of the helium bath or due to the microphonics. In addition to compensating for the rapid detuning induced by the Lorentz force, Fermilab's adaptive LFD algorithm monitors the cavity resonance frequency and adaptively adjusts the DC bias on the piezo actuator to stabilize the resonance frequency. The resonant frequency of the CM2 cavities over a period of several hours drift by up to 20Hz when adaptive compensation is not active. When adaptive compensation is on the drift was suppressed to less than one Hz.

An adaptive LFD compensation system was successfully used to compensate for large (up to 1000Hz) Lorentz force detuning in all 8 cavities in CM2 during commissioning and operation.

## 3. Record high gradient beam acceleration

The FAST beamline starts with a photocathode in a normal-conducting electron RF gun with 1.5-cell copper cavity operating at 1.3 GHz – see Fig.2. A train of UV pulses from the drive laser strikes a $Cs_2Te$-coated Mo cathode in the gun cavity resulting in a train of 4.5 MeV electron bunches with the bunch charge of up to 3.2 nC. The pulse train structure is selectable between a single pulse for the IOTA ring injection beam cycle (nominally 1 Hz) and up to a 1 ms long trains of 3000 bunches 333ns apart and operates with repetition rate up to 5 Hz – see Fig.6. The beam then passes through a short (~1 m) low-energy diagnostic section before acceleration in two consecutive, individually powered superconducting RF accelerating structures referred to as the "capture cavities", CC1 and CC2. Each capture cavity in its own cryostat is a 9-cell, 1.3 GHz Nb accelerating structure nominally cooled to 2°K. Following acceleration of up to maximum 52 MeV, the electron beam passes through the low-energy beam transport section, which includes steering and focusing elements, an optional chicane for bunch compression and beam transforms, as well as a set of beam diagnostics instrumentation. It is then either directed into the low-energy absorber by the D122 dipole magnet or goes to the CM2, which further accelerates the beam to about 300 MeV beam. The high-energy beam is transported through a long beamline either to a high energy beam absorber or, in the future, into the IOTA ring injection section. Table 3 presents main parameters of the linac beam. The FAST beamline is well equipped with necessary types of beam diagnostics (see details in [15]): beam position monitors (BPMs), transverse profile monitors to measure beam size, resistive wall current monitors to measure beam current and loss monitors to measure beam losses and serve as the primary protection element in the machine protection system, while the beam energy is measured with use of two spectrometer magnets of D122 and D600.

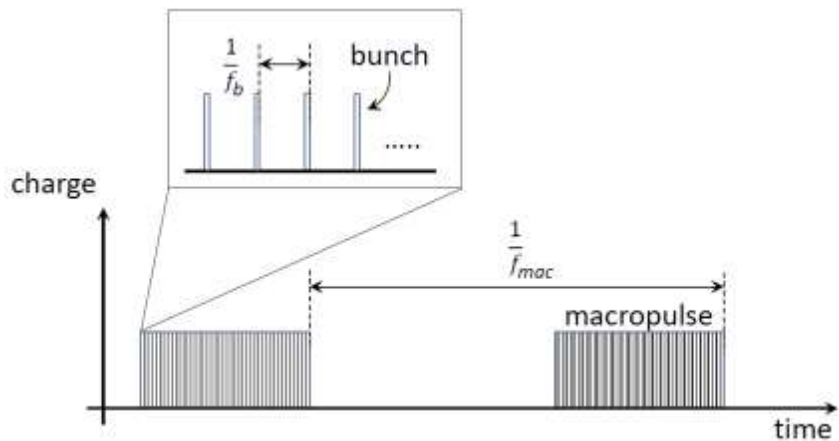

Figure 6: Time structure of the IOTA/FAST electron injector beam. $f_{mac}$ is nominally 1-5 Hz, macropulse length is 0.1-1ms, and $f_b$ is up to 3 MHz.

Table 3: Main beam parameters of the IOTA/FAST electron injector.

| Parameter | FAST Value |
|---|---|
| Beam energy to low energy absorber | 20 – 52 MeV |
| Beam energy to high energy absorber | 100 – 300 MeV |
| Bunch charge, $q$ | < 10 fC – 3.2 nC |
| Bunch frequency, $f_b$ | 0.5 – 9 MHz |
| Micropulse frequency, $f_{mac}$ | 1 – 5 Hz |
| Micropulse duration | < 1 ms |
| Normalized emittance, $\varepsilon_n$ (at 0.1nC/bunch) | 0.6 mm mrad |

*3.1 Cryomodule tuning*

The optimization of RF phases between the gun, CC1 and CC2 is fairly simple, given that each has its own low-level RF (LLRF) and high level RF systems, and maximizing the total energy gain is relatively easy with only three cavities. This is not the case for the CM2, in which 8 cavities are powered by one RF source and controlled by a common low-level RF system, resulting in a vector sum of individual cavity energy gains $U_i$

$$\Delta W = \sum_{i=1}^{8} U_i \cos(\varphi_i) \quad (1)$$

that can rapidly become difficult to deconvolve as each relative phase and amplitude change could impact the remaining cavities. To find the optimal position for the individual cavity phase $\varphi_i$ tuners, we employed a beam-loading-based tuning method. Beam loading is the effect of interaction of the beam current with the cavity – if the beam is accelerated by the cavity, it results in the increase of the beam energy and corresponding reduction of the energy of electro-magnetic fields stored in the cavity; if the beam is out of phase with cavity's RF field it gets decelerated and therefore its energy is transferred to the cavity. For a short bunch with charge $q$, and the amplitude of the beam-induced wakefield in the cavity is given by [30]:

$$\Delta U = q \left(\frac{R}{Q}\right) \frac{\omega}{2} \quad (2)$$

where *(R/Q)* is the cavity geometry dependent parameter of about 1000 Ohms for the ILC-type cavities, and $\omega$ is the frequency of the exited mode, $2\pi \times 1.3$ GHz in our case. Cavity monitors of the RF field thus would report the variation of the total RF voltage amplitude depending on relative phase of the RF and the beam $\Delta\varphi=\varphi_i-\varphi_b$. Adjustment of either the low-level RF phase or the phase delay tuning motor could be used to shift the RF phase change corresponding to the beam loading. A phase shift over the complete RF cycle will include both maximum deceleration and maximum acceleration through the cavity, the latter corresponding to the maximum beam loading, as shown in Fig.7 for cavity 1. Choosing to maximize the acceleration for all cavities to a LLRF phase set-point of 0, the phase delay tuning motor was adjusted to maximize the peak beam loading in the cavity.

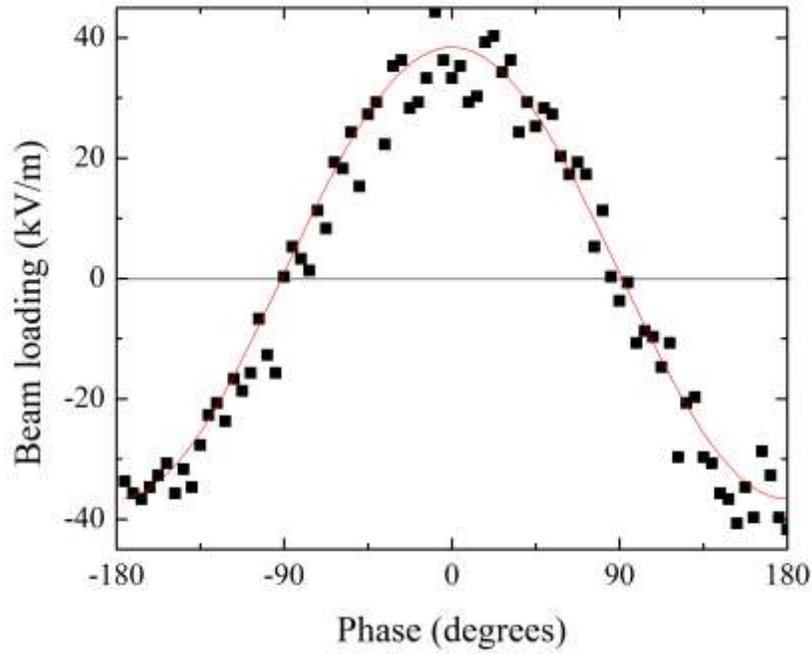

Figure 7: Beam loading effect (reduction of the 1.3 GHz RF signal amplitude) in cavity 1 vs the RF phase.

This process was repeated for the remaining seven cavities and optimal phases were determined and set for all of them. Once all eight CM2 phase delay tuners were optimized, the beam was accelerated down the high energy beamline to the energy of 298.0±3.4 MeV measured after the 15-degree dipole D600, which was used as the high energy spectrometer. There are two sources of errors in the energy measurement. First, it is the uncertainty in the beam angles before and after the dipole, measured by corresponding beam-position monitors (BPMs), which is estimated to be about 0.9% (2 mrad for the total bend angle of 15 degrees). The second is the uncertainty of magnetic field calibration of the D600 magnet at the level of 0.7% rms. Stability of the energy past CM2 even with open-loop LLRF stabilization system was better than 0.1% rms, as derived from the beam position jitter in the BPM past D600.

The final beam energy is composed of two parts gained in the low energy section before CM2 and in the CM2. The first part was precisely measured with spectrometer magnet D122. That magnet was calibrated and had the NMR probe in it to measure actual magnetic field. Two pairs of BPMs are located immediately upstream and downstream of the D122 to measure the change of the beam trajectory angle. The resulting injector beam energy was found to be 42.9±0.2 MeV for that particular run, therefore, the CM2 total acceleration was 255.1 ± 3.4 MeV.

*3.2 Lattice tuning*

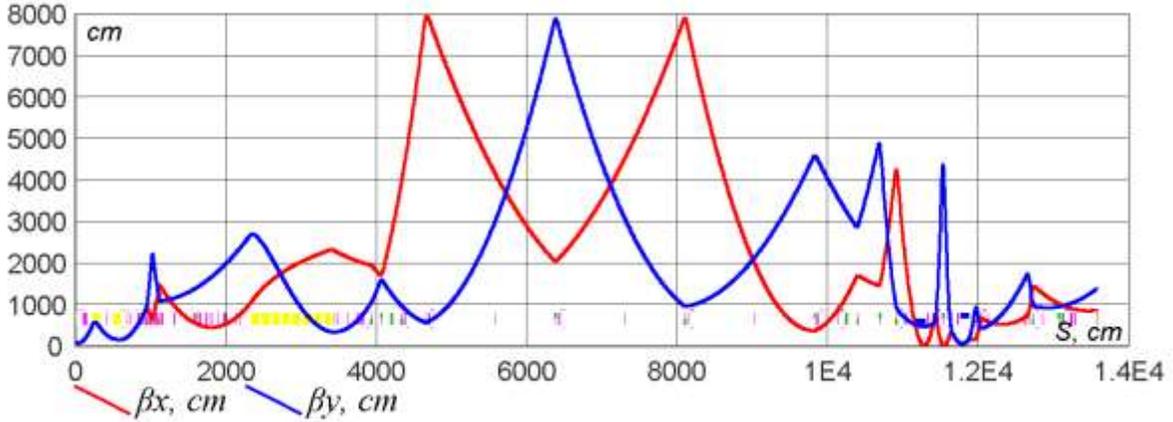

Figure 8: Optical functions $\beta_{x,y}(s)$ along the 300 MeV IOTA/FAST electron injector linear accelerator. The origin is at the photo-injector cathode, and the beam ends at the high energy absorber. Presented are design values, measured beta-functions are commonly within ± 10% from the design.

Careful control of the beam size and trajectory is needed to assure lossless acceleration and transport of electron bunches from the gun to the high energy absorber, and, in the future, deliver suitable beam for the injection into the IOTA ring. Beam dynamics in a linear accelerator is defined by the initial conditions at a starting point and by the external perturbations in the downstream lattice. In a simplified form, the beam trajectory *x(s)* deviates from a perfect, e.g. straight, line, due to transverse angular kicks $\theta_k$ caused by imperfections, such as misalignment of magnets, RF cavities, etc in the preceding sections of the beamline [31]:

$$x(s) = \sum_k \theta_k \sqrt{\beta(s)\beta_k} \times \sin(\varphi(s) - \varphi_k) \quad (3)$$

while the rms beam size varies as:

$$\sigma(s) = \sqrt{\beta(s)\varepsilon_n/\gamma} \quad (4)$$

Here, *β(s)* is beta-function of the focusing lattice at the location of observation *s* – see Fig.8 – and $\beta_k$ are its value at the locations of the kicks, $\varepsilon_n$ is the normalized beam emittance determined mostly by the electron source, $\theta_k$ is the dipole kick due to the *k*-th imperfection, and the betatron oscillation phase $\varphi(s)=\int ds/\beta(s)$. After the initial manual tuning of propagation of the beam to the absorber, alternating steps of beam-based focusing lattice and trajectory corrections were made. The trajectory correction was aimed at minimizing beam offsets from the magnetic axis of quadrupoles. The analysis and correction of the focusing optics was done using the Linear Optics from Closed Orbit (LOCO) technique [32, 33]. Operationally important aspects of the focusing lattice tuning were optimization of the beam shape distortions at low energies (4.5 to 20 MeV) to minimize space-charge induced degradation of emittances at high bunch charge [34], careful control of the orbits through slightly misaligned capture cavities CC1 and CC2 and proper correction of the dispersion achromat in the dogleg formed by the two 15-degree dipoles (at around 110-120m in Figure 8) [35]. Figure 9 present the beam profile images on the transverse profile monitor downstream of the D600 dipole before and after the beamline tuning showing a great deal of improvement resulting in loss-free transportation of optimal emittance beam.

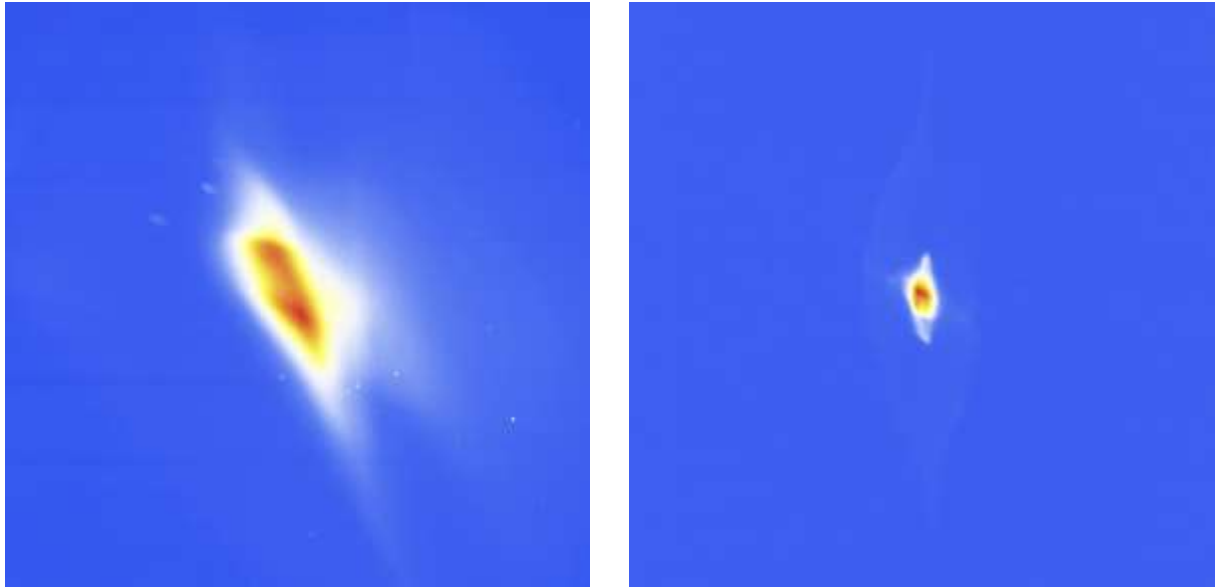

Figure 9: High energy electron beam images from the beam after D600 as observed on the YAG:Ce crystal screen of the FAST transverse profile monitor: (left) before and (right) after the orbit correction and beamline focusing lattice tuning. Vertical and horizontal screen sizes are 20mm.

## 4. Conclusion

We have commissioned 1.3 GHz superconducting RF linac at the Fermilab Accelerator Science and Technology facility and have demonstrated record high beam accelerating gradient ever achieved in large-scale SRF accelerators. Energy gain in the eight cavity 1.3 GHz SRF cryomodule CM2 exceeded 255 MeV and the average beam accelerating gradient matched the ILC specification of 31.5MV/m. Attainment of such record gradients is the result of more than a decade of R&D and was possible due to a) advances in the technology of the SRF cavities production, surface treatment and cryomodule assembly; b) development of sophisticated low-lever RF system and Lorentz force detuning compensation system to control and stabilize cavities frequencies, gradients and phases at the very low levels required for efficient beam acceleration; and c) construction and commissioning of a modern electron photo-injector and high-energy beam transport line equipped with efficient beam diagnostics and control system to allow efficient tuning of the SRF accelerator parameters as well as beam trajectory, focusing optics and emittances.

The FAST facility 2017 operation with two eight-hour shifts per day allowed to combine active cryomodule and beamline commissioning and tuning with several beam experiments carried out in collaboration with external and internal research groups, such as studies of effects of the high order modes (HOM) excited by the beam in CC1 and CC2 on the transverse beam emittance [36], tests of the advanced beam diagnostics using synchrotron radiation from the D600 dipole [37], experimental verification of machine learning algorithms for optimization of the low-energy accelerator injector [38], investigation of the 4D beam phase space tomography [39] and innovative experiments on the round-to-flat beam transformations of magnetized high intensity electron beams [40, 41].

The achieved beam parameters – energy, intensity, stability and emittance – are fully adequate to the specifications of the program of accelerator R&D at the IOTA ring toward intensity frontier beams [15, 42]. Besides this main goal, the facility is capable to support a broad spectrum of advanced beam studies and experiments [43]. In particular, it will be quite suitable for the beam tests of a number of novel methods to further significant increase of the accelerating gradients up to 90MV/m in pulsed SRF cavities, such as those now actively developed, using of $Nb_3Sn$ rather than pure Nb cavities [44, 45, 46] or based on impurity addition to the cavity surface or layered structures of insulating and superconducting films [47, 48].

## *Acknowledgements*

This report is a climax of decadal work of numerous expert groups from Fermilab's Accelerator and Technical Divisions on fabrication and construction of the SRF accelerator and the FAST facility, its cryogenics, vacuum, beam diagnostics and control system and many other critical components. We sincerely thank them for their accomplishments. We particularly acknowledge useful discussions with S.Posen, A.Grasselino and A.Romanenko on the prospects of various approaches toward very high SRF accelerating gradients.

Fermilab is operated by Fermi Research Alliance, LLC under Contract No. DE-AC02-07CH11359 with the United States Department of Energy.